\documentclass[prl,aps,twocolumn,superscriptaddress]{revtex4-1}
\usepackage{amssymb,amsfonts,amsmath,graphicx,color,mathptm,times}

\def\tr{\mbox{tr}}
\def\bra#1{\langle{#1}|}
\def\ket#1{|{#1}\rangle}
\def\braket#1{\langle{#1}\rangle}

{\catcode`\|=\active 
  \gdef\Braket#1{\begingroup
\mathcode`\|32768\let|\BraVert\left<{#1}\right>\endgroup}}
\def\BraVert{\egroup\,\mid\,\bgroup}

\newcommand{\q}{\mathbf{Q}}

\allowdisplaybreaks

\begin{document}

\title{Measuring the heat exchange of a quantum process}

\author{John Goold}
\email{jgoold@ictp.it}
\affiliation{The Abdus Salam International Centre for Theoretical Physics (ICTP), Trieste, Italy}

\author{Ulrich Poschinger}
\email{poschin@uni-mainz.de}
\affiliation{QUANTUM, Institut f\"{u}r Physik, Universit\"{a}t Mainz, Staudingerweg 7, 55128 Mainz,
Germany}

\author{Kavan Modi}
\email{kavan.modi@monash.edu}
\affiliation{School of Physics, Monash University, VIC 3800, Australia}

\date{\today}

\begin{abstract}
Very recently, interferometric methods have been proposed to measure the full statistics of work performed on a driven quantum system [R. Dorner {\it et al.} {\it Phys. Rev. Lett.} {\bf 110}, 230601 (2013) and Mazzola {\it et al.} {\it Phys. Rev. Lett.} {\bf 110}, 230602 (2013)]. The advantage of such schemes is that they replace the necessity to make projective measurements by performing phase estimation on an appropriately coupled ancilla qubit. These proposals are one possible route to the tangible experimental exploration of quantum thermodynamics, a subject which is the center of much current attention due to the current control of mesoscopic quantum systems. In this Rapid Communication we demonstrate that a modification of the phase estimation protocols can be used in order to measure the heat distribution of a quantum process. In addition we demonstrate how our scheme maybe implemented using ion trap technology.  Our scheme should pave the way for the first experimental explorations of the Landauer principle and hence the intricate energy to information conversion in mesoscopic quantum systems.  
\end{abstract}

\maketitle

{\it Introduction.---} Landauer's principle states that the heat generation in an irreversible computation must always be greater than or equal to the information theoretic entropy change~\cite{landauer}. The result is undoubtedly one of the deepest results of modern day computer science and information theory, providing a definitive link between energy and information. So profound is the principle that Bennett used it in order to exorcise Maxwell's demon by attributing a minimum entropy production to the logically irreversible procedure of erasure~\cite{bennett}. 

It is indeed surprising that, despite its simplicity, the Landauer principle has only just been verified experimentally~\cite{lutz, orlov}. In this experiment the mean heat of a single colloidal particle trapped in a double well-potential was measured. Performing the requisite erasure procedure by modulating the double well, the average dissipated heat was found to saturate the Landauer bound in the long time limit. 

Turning to quantum systems, experiments in this direction still need to be performed. Of course, the Landauer principle is expected to hold generally, irrespective of the underlying classical or quantum nature of the system. However, recent work by Reeb and Wolf~\cite{reeb} has demonstrated that, for finite--dimensional quantum systems, the Landauer principle can be tighter by an amount which depends on the size of the thermal reservoir. 

Undoubtedly, any experiment which aims at exploring the fundamental energetic limits of information processing would need to measure the {\it heat} exchange in a fundamental process. The modern approach to the thermodynamics of small systems is the framework of stochastic energetics~\cite{sekimoto} whereby quantities such as heat and work are described by probability distributions. These distributions obey fluctuation relations which have been extensively explored, both theoretically and experimentally, since their inception~\cite{jrev}. The fluctuation relations, extended to the quantum mechanical domain~\cite{mrev, Tasaki, esposito}, are a promising route to understand the statistical physics of small quantum systems which are operating under nonequilibrium conditions. However, due to the additional fragility of quantum systems, the experimental extraction of the relevant distributions has been hampered. Recent work has demonstrated that quantum ``work" statistics maybe extracted by means of quantum tomography of a coupled ancilla~\cite{dorner2, mauro}. This theoretical work has paved the way to the first experimental demonstration of the quantum work fluctuation relations in a liquid state NMR setup~\cite{nmr}. For possible extensions to strongly coupled open systems see~\cite{openp, openmauro, open}.

In this Rapid Communication we demonstrate that a phase estimation scheme conceived in a similar spirit to~\cite{dorner2, mauro} maybe used in order measure out the characteristic function and hence the probability distribution of heat in a generic quantum process. The first moment of the distribution is the average heat and maybe used to explore the Landauer principle and information to energy conversion in the quantum domain~\cite{refsug}.  We demonstrate the feasibility of our proposal using realistic parameters for the example of two trapped calcium ions interacting with external laser fields.

{\it Setting.---} Consider a system ($S$) on which we would like to perform a protocol with the aid of a finite dimensional reservoir ($R$) (it could be any computation, such as erasure or a work extraction process). We assume that initially the total ($RS$) state has no correlations: $\rho_{RS}=\rho_R \otimes \rho_S$. We additionally assume that the initial state of the reservoir $\rho_R$ is of Gibbs form:
\begin{gather}\label{gibbs}
\rho_R = \sum_m \frac{e^{-\beta E_m}}{Z_R} \ket{r_m}\bra{r_m},
\end{gather}
where $Z_R = \sum_m e^{-\beta E_m}$ with $H_R = \sum_m E_m \ket{r_m}\bra{r_m}$, and $\beta$ is the inverse temperature. Now we perform a global unitary (the protocol) on the composite ($RS$) state: 
\begin{gather}
\rho_{RS} \to \rho'_{RS} = U \rho_R \otimes \rho_S U^\dag
\end{gather} 
and $\rho'_R = \tr_S[\rho'_{RS}]$ and $\rho'_S = \tr_R[\rho'_{RS}]$. In order to assume no net work has been done we assume the Hamiltonian of the reservoir is described by a fixed $H_R$. Under this set of assumptions the change of the energy in the reservoir is the average heat of the process~\cite{reeb},
\begin{gather}\label{avheat}
\braket{\q} = \tr[H_R \, \rho'_R]- \tr [H_R \, \rho_R].
\end{gather}
However, care must be taken in interpreting this quantity as heat in
the most general sense because in the strong-coupling regime the
division of the energy changes into heat and work becomes unclear. However, one can still define an energy dissipation to the reservoir in order to avoid any issues of interpretation.  

\begin{figure}
\includegraphics[width=\linewidth]{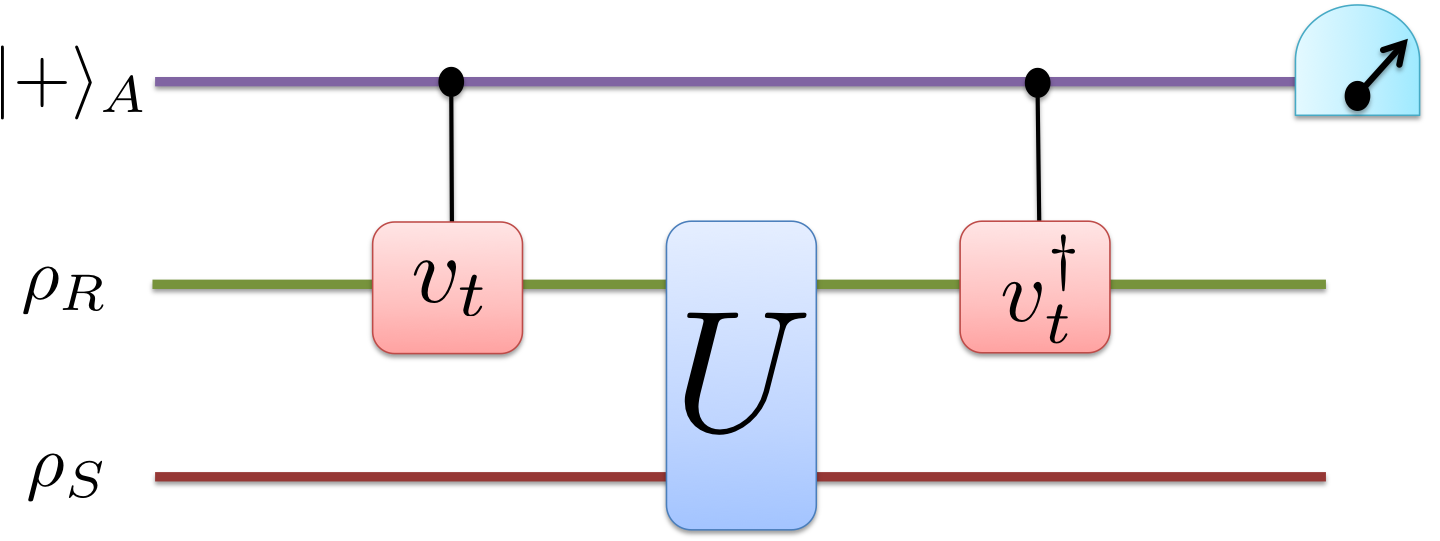}
\caption{(Color online.) The quantum circuit which is used to measure the heat of a quantum process. The ancilla qubit in the upper branch is prepared in a $\ket{+}$ state, the system of interest is prepared in an arbitrary initial state, whereas the reservoir state $\rho_R$ defined in the text is a thermal state. First, a controlled operation $v^{\dagger}=e^{i H_{R}t}$ is applied on the reservoir, next, the protocol unitary $U$ is applied, and then another controlled operation $v$ is performed on the reservoir and the qubit is measured in the $x-y$ plane. \label{fig1}}
\end{figure}

It is important to point out that if we are dealing with a truly microscopic system both quantum and thermal fluctuations will be prominent \cite{talknerheat}. In fact the heat exchanged is actually the first moment of a total probability distribution for heat $P(\q)$,
\begin{gather} \label{heatdist}
 P(\q)=\sum_{mn}p_{m}p_{n|m}\delta(\q-(E_{n}-E_{m})).
\end{gather}
This distribution is equivalent to the marginal distribution of the joint distribution studied in \cite{talknerheat} and it is important to stress it is in fact that it is only the joint distribution which satisfies a fluctuation relation of the standard form \cite{mrev}. The distribution is built by the following procedure: Before the unitary protocol is applied the reservoir is projectively measured to have energy $E_{m}$ with probability given by the Boltzmann factor $p_{m}={e^{-\beta E_m}}/{Z_R}$, then a generally non unitary dynamics occurs on the reservoir (and the system) and the energy is measured again with conditional probability $p_{m|n}=\tr[U \ket{r_{m}} \bra{r_{m}} \otimes \rho_{S} \, U^{\dagger} \ket{r_{n}} \bra{r_{n}}]$ thus forming a distribution of heat changes. It is important to stress that the dynamics of the reservoir is not unitary and the problem may have been set up from the beginning using the approach of describing the reservoir (system) dynamics using completely positive and trace preserving maps. This approach has recently been taken in order to derive fluctuation like relations for general quantum
channels \cite{kafri,vedral,rastegin,albash,rastegin2,recent}. In \cite{kafri,vedral,rastegin,albash,rastegin2,recent} the relationship between the non-unitality of a channel and the microreversibility of
the process was studied, which is indeed an interesting link between
the non-unitality of a channel and a bound on the heat dissipated in a generic  quantum process \cite{bound}.

{\it Measuring the heat distribution.---} The heat distribution Eq.~\eqref{heatdist} has a corresponding characteristic function or cumulant generating function defined by a Fourier transform
\begin{gather}
\Theta(t)=\int_{-\infty}^{\infty} P(\q) \, e^{i t \q} \, \textrm{d} \q,
\end{gather}
carrying out the Fourier transform we can recast $\Theta(t)$ in the following compact form
\begin{align}
\Theta(t)=& \sum_{mn}p_{m}p_{n|m}e^{-i(E_{n}-E_{m})t}
\nonumber \\
=& \sum_{lmn} \frac{e^{-\beta E_{l}}}{Z_{R}}e^{-i (E_{m}-E_{n})t} \notag\\
& \times \tr[U \ket{r_{l}}\braket{r_{l} | r_{m}} \bra{r_{m}} \otimes \rho_{S} \, U^{\dagger} \ket{r_{n}} \bra{r_{n}}] \notag \\
=& \textrm{tr}[U \, \rho_R \, v^{\dagger} \otimes \rho_S \, U^{\dagger} \, v],
\label{charfunction}
\end{align}
with the new unitary operator $v_t=e^{-i H_{R} t}$. We stress that the first cumulant in an expansion of Eq.~\eqref{charfunction} corresponds to average heat defined by Eq.~\eqref{avheat}.

Consider the quantum circuit in displayed in Fig.~\ref{fig1}. An ancilla qubit ($A$) $\rho_{A}$ is brought in contact with our system ($S$) and reservoir ($R$) (in fact, $A$, $S$, and $R$  can all be qubits as we made no restrictions on the dimension of either the system or the reservoir). Let us label the total state in the $k$th step as $\rho^{(k)}_{ARS}$ and go through the steps of the interferometer.

The ancilla is prepared initially in the $\ket{+_{A}} = (\ket{0_{A}} + \ket{1_{A}})/\sqrt{2}$ state. The initial total state is
\begin{align}
\rho^{(0)}_{ARS} = \frac{1}{2} & \left(
   \ket{0_{A}}\bra{0_{A}} \otimes \rho_R \otimes \rho_S
 + \ket{0_{A}}\bra{1_{A}} \otimes \rho_R \otimes \rho_S \right. \notag\\
&+ \left. \ket{1_{A}}\bra{0_{A}} \otimes \rho_R \otimes \rho_S 
 + \ket{1_{A}}\bra{1_{A}}  \otimes \rho_R \otimes \rho_S \right).
\end{align}
We can restate the last equation in a more compact form by writing it as a matrix in the basis of $A$
\begin{gather}
\rho^{(0)}_{ARS}= \frac{1}{2} \left(\begin{matrix} 
\rho_R \otimes \rho_S & \rho_R \otimes \rho_S \\
\rho_R \otimes \rho_S & \rho_R \otimes \rho_S \\
\end{matrix}\right).
\end{gather}
In the next step, the unitary operation $v_t=e^{-iH_{R}t}$ is applied on the reservoir when $A$ is in state $\ket{1_{A}}$ (controlled-operation) yielding the total state
\begin{gather}
\rho^{(1)}_{ARS} = \frac{1}{2} \left(\begin{matrix} 
\rho_R \otimes \rho_S & \rho_R \, v_t^\dag \otimes \rho_S \\
v_t \, \rho_R \otimes \rho_S  & v_t \, \rho_R \, v_t^\dag \otimes \rho_S \\
\end{matrix}\right).
\end{gather}
Next, the unitary protocol $U$, whose energetics we wish to investigate, is applied on $RS$, yielding the total state
\begin{gather}
\rho^{(2)}_{ARS} = \frac{1}{2} \left(\begin{matrix} 
U \, \rho_R \otimes \rho_S \, U^\dag  
& U \, \rho_R \, v_t^\dag \otimes \rho_S \, U^\dag \\
U \, v_t \, \rho_R \otimes \rho_S \, U^\dag 
& U \, v_t \, \rho_R \, v_t^\dag \otimes \rho_S \, U^\dag \\
\end{matrix}\right).
\end{gather}
Finally the second controlled unitary transformation $v_t^\dag$ is applied on $R$ to give 
\begin{gather}
\rho^{(3)}_{ARS} = \frac{1}{2} \left(\begin{matrix} 
U \, \rho_R \otimes \rho_S \, U^\dag  
& U \, \rho_R \, v_t^\dag \otimes \rho_S \, U^\dag \, v_t \\
v_t^\dag \,U \, v_t \, \rho_R \otimes \rho_S \, U^\dag
& v_t^\dag \, U \, v_t \, \rho_R \, v_t^\dag \otimes \rho_S \, U^\dag \, v_t \\
\end{matrix}\right)\notag.
\end{gather}

The state of $A$ is obtained by tracing over $RS$:
\begin{gather}
\rho^{(3)}_{A} = \frac{1}{2} \left(\begin{matrix} 
1  & \tr[U \, \rho_R \, v_t^\dag \otimes \rho_S \, U^\dag \, v_t] \\
\tr[v_t^\dag \,U \, v_t \, \rho_R \otimes \rho_S \, U^\dag] & 1 \\
\end{matrix}\right)\notag.
\end{gather}
$A$ is now measured in the $x-y$ plane yielding access to the characteristic function Eq.~\eqref{charfunction}:
\begin{gather}
\Theta(t) = \tr[(X_A - i Y_A) \rho_A^{(3)}],
\end{gather}
where $X_A$ and $Y_A$ are Pauli operators on the space of the ancilla. The heat distribution and its moments may then be extracted via an anti-Fourier transform of this signal~\cite{nmr}. It is also worth pointing out that, strictly speaking, the necessity to keep the reservoir in the Gibbs state Eq.~\eqref{gibbs} maybe relaxed in favor of a so called {\it passive} state~\cite{polish} of which the thermal state is a particular case.  

\begin{figure}[t]
\includegraphics[width=0.5\textwidth]{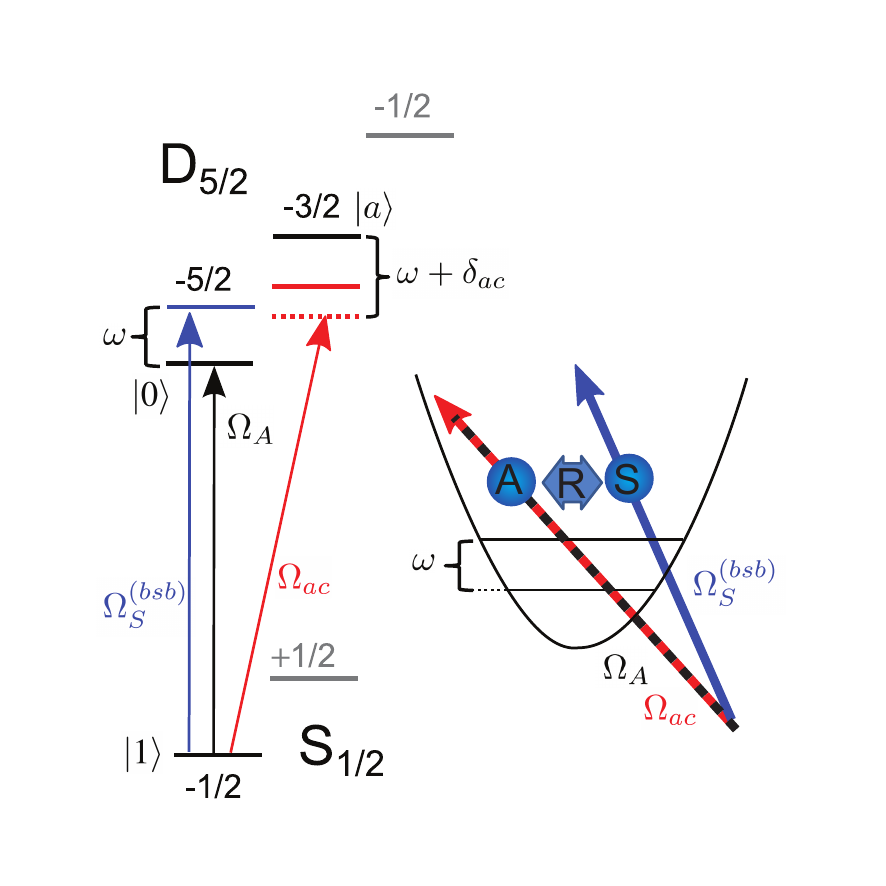}
\caption{(Color online.) Proposal to measure the heat distribution with trapped ions. The level scheme shows the sublevels of $^{40}$Ca$^+$ ions for encoding the system and ancilla qubits, along with the required laser fields for the qubit state manipulation, system-reservoir interaction. On the right, two ions in a common harmonic trap are shown along with the different laser beams. \label{fig:experiment}}
\end{figure}

{\it An experimental proposal with trapped ions.---} We propose an implementation of our scheme based on laser-cooled trapped ions. We consider two $^{40}$Ca$^+$ ions of mass $m$ confined in a harmonic potential. The setting and relevant levels are indicated in Fig.~\ref{fig:experiment}. The qubits $S$ and $A$ which are needed for the Landauer circuit of Fig.~\ref{fig1} are encoded in the states $\ket{0} \equiv \ket{D_{5/2},m_J=-5/2}$ and $\ket{1} \equiv \ket{S_{1/2},m_J=-1/2}$ of the different ions. The $D_{5/2}$ state is metastable with a lifetime of about 1s. An external quantizing magnetic field gives rise to a Zeeman splitting of the $m_J$ sublevels, which is typically in the range of 2$\pi\times$5 to 2$\pi\times$20MHz. Qubit rotations can be driven by means of resonant laser irradiation near $\lambda=$729nm. A normal mode of vibration at frequency $\omega$, typically in the range of 2$\pi\times$1 to 2$\pi\times$5MHz, acts as the reservoir $R$, and it can be conveniently initialized to a Gibbs state, as required in the protocol, by Doppler cooling and resolved sideband cooling~\cite{u1}. The temperature can be varied from 2~mK to below 6~$\mu$K. The laser-induced coupling between the qubits and the reservoir is characterized by the Lamb-Dicke parameter $\eta = 2 \pi \cos(\theta) \sqrt{\hbar/(2m\omega)} /\lambda$, where $\theta$ is the laser angle of incidence with respect to the oscillation direction.  The laser can either be addressed to the ions by controlling its propagation direction~\cite{u1}, or alternatively in frequency space by employing a strong magnetic field gradient~\cite{u2} or spatially inhomogeneous dressing fields~\cite{u3}.\\

The RS unitary can be generated by resonantly driving the blue sideband transition $\ket{1_S,n}\leftrightarrow \ket{0_S,n+1}$ of the $S$ qubit at Rabi frequency $\Omega_S^{(bsb)}\approx \eta\sqrt{n+1}\Omega_S$. The corresponding Hamilton operator reads
\begin{gather}
H_{RS}=\frac{1}{2} \Omega_{S} \left(\ket{0_S}\bra{1_S}\otimes b_R^{\dagger} +\ket{1_S}\bra{0_S}\otimes b_R \right).
\end{gather}
The controlled Ancilla-Reservoir interaction $\nu_t$ is implemented by means of a \textit{quantized ac-Stark shift}~\cite{u4}: A motional sideband is driven off-resonantly, such that a phase shift proportional to the phonon number is obtained. To render this phase shift conditional on the state of $A$, we drive the transition to an auxiliary level $\ket{a} \equiv \ket{D_{5/2},m_J=-3/2}$ The Hamilton operator reads
\begin{align}
H_{AR} =& \frac{1}{2} \Omega_{ac} \left(\ket{a_A}\bra{1_A}\otimes b_R+\ket{1_A}\bra{a_A}\otimes b_R^{\dagger} \right) \notag\\ 
&+ \frac{1}{2} \delta_{ac}\left(\ket{a_A}\bra{a_A}-\ket{1_A}\bra{1_A}\right) \otimes \openone_R \notag\\
\simeq&\frac{\eta^2\Omega_{ac}^2}{4 \delta_{ac}}\ket{1_A}\bra{1_A} \otimes N_R,
\end{align}
where $\delta_{ac}$ is the detuning from the red motional sideband, $\Omega_{ac}$ is the driving strength of the $\ket{1_A}\leftrightarrow\ket{a_A}$ carrier transition, $b_R(b_R^{\dagger})$ are the reservoir annihilation(creation) operators and $N_R=b_R^{\dagger}b_R$ is the reservoir number operator. In the second line, we use $\delta_{ac}\gg \eta\Omega_{ac}$ to adiabatically eliminate the $\ket{a}$ state. For a drive time $t$, this leads to the unitary 
\begin{gather}
\nu_t= \ket{0_A}\bra{0_A} \otimes \openone_R+
\ket{1_A}\bra{1_A} \otimes
\exp\left(-i\frac{\eta^2\Omega_{ac}^2}{4 \delta_{ac}} N_R t\right).
\end{gather}
In order to generate the adjoint operation $\nu_t^{\dagger}$, the sign of $\delta_{ac}$ has to be reversed. For $\eta=$0.07, $\delta_{ac}=2\pi \times$~100~kHz and $\Omega_{ac}=2\pi \times$~300~kHz, a conditional frequency shift of $2\pi\times$~1.1~kHz per phonon is achieved, while  less than 3\%~$+\sqrt{n}\times$~5\% of the population is cycled through $\ket{a}$. 

Estimates for the diagonal elements of the  density matrix $\rho_A$ are obtained by repeating a measurement for constant parameters a sufficiently large number of times $M$, where the statistical error is scaling as $1/\sqrt{M}$. The real and imaginary parts of a given value of the characteristic function, Eq.~\eqref{charfunction}, are read out by modulating the phase of the second $\pi/2$ pulse on $A$. This phase can be controlled e.g. with an acousto-optical modulator. For $\phi=0$, the probability to measure $A$ in $\ket{0_A}$ is $(1+\text{Im}\Theta)/2$, while for $\phi=\pi/2$, it is $(1+\text{Re}\Theta)/2$. 

{\it Conclusions.---} In this Rapid Communication we have outlined a clear and straightforward interferometric scheme for the measurement of the heat of a quantum process. Our scheme is not restricted to quasi-static protocols and the full statistics of the quantum and thermal fluctuations maybe studied by means of an ancillary system. Given the success of the schemes~\cite{dorner2, mauro} in bringing forth the first experimental extraction of quantum work statistics~\cite{nmr} in a Liquid state NMR setup, we believe that the scheme presented here will provide inspiration for the first generation of experiments to test the thermodynamics of computational protocols operating deeply in the quantum regime. In its current form, our scheme is readily implementable using a variety of different experimental platforms such as trapped ions, as demonstrated here. One may also hope that the formalism outlined here may be extended to measure heat dissipated in many-body systems following various quench protocols, where interesting links with critical features are currently been explored \cite {latent}. Most importantly, we hope that our proposal will inspire the first experimental explorations of the relationship between energy and information in the quantum domain. 

\begin{acknowledgments}
{\it Acknowledgments.---} The authors thank Lucas Celeri for his hospitality at the Federal University of Goi\'as. We thank Lucas Celeri, Nicolino Lo Gullo, Du Du ``{\it that's my point}" Mascarenhas, Mauro Paternostro, and Roberto Silva Sarthour for insightful discussions. KM was supported by the John Templeton Foundation, the National Research Foundation, and the Ministry of Education of Singapore during completion of this work. Part of this work was supported by the COST Action MP1209 ``Thermodynamics in the quantum regime".
\end{acknowledgments}



\begin{thebibliography}{99}

\bibitem{landauer} R. Landauer, IBM J. Res. Dev. {\bf 5}, 183 (1961).

\bibitem{bennett} C. H. Bennett, IBM J. Res. Dev. {\bf 17}, 525-531 (1973); C. H. Bennett, Int. J. Theor. Phys. {\bf 21}, 905-940 (1982); C. H. Bennett, St. Hist. \& Phil. of Mod. Phys., {\bf 34}, 501-510 (2003).

\bibitem{lutz} A. Berut, A. Arkelyan, A. Petrosyan, S. Ciliberto, R. Dillenschneider and E. Lutz, Nature {\bf 483} 187-189, (2012).

\bibitem{orlov} A. O. Orlov, C. S. Lent, C. C. Thorpe, G. P. Boechler and G. L. Snider, Jpn. J. Appl. Phys. {\bf 51}, 06FE10 (2012).

\bibitem{reeb} D. Reeb and M. W. Wolf, {arXiv:1306.4352} (2013).

\bibitem{sekimoto} K. Sekimoto, {\it Stochastic Energetics} (Springer, Lecture notes in physics Vol. 799, (2010)).

\bibitem{jrev} C. Jarzynski, Annu. Rev. Cond. Matter Phys. {\bf 3}, 329 (2011).

\bibitem{mrev} M. Campisi, P. H\"{a}nggi and P. Talkner, Rev. Mod. Phys. {\bf 83}, 771 (2011).

\bibitem{Tasaki} H. Tasaki, {arXiv:cond-mat/0009244} (2000);  J. Kurchan, {arXiv:cond-mat/0007360v2} (2000); S. Mukamel, Phys. Rev. Lett. {\bf 90}, 170604 (2003).

\bibitem{esposito} M. Esposito, U. Harbola and S. Mukamel, Rev. Mod. Phys. {\bf 81}, 1665 (2009).

\bibitem{dorner2} R. Dorner, S. R. Clark, L. Heaney, R. Fazio, J. Goold, and V. Vedral (2013) Phys. Rev. Lett.  {\bf 110} 230601.

\bibitem{mauro} L. Mazzola, G. De~Chiara, and M. Paternostro (2013)  Phys. Rev. Lett. {\bf 110} 230601.

\bibitem{nmr} T. Batalhao, A. M. Souza, L. Mazzola, R. Auccaise, I. S. Oliveira, J. Goold, G. De~Chiara, M. Paternostro and R. M. Serra, {arXiv:1308.3241} (2013).

\bibitem{openp} M. Campisi, R. Blattman, S. Kholer, D. Zueco, and P. H\"{a}nggi, New J. Phys. {\bf 15}, 105028 (2013).

\bibitem{openmauro} L. Mazzola, G. De~Chiara, and M. Paternostro, arXiv:1401.0566 (2014).

\bibitem{open} M. Campisi, P. Talkner, and P. H\"{a}nggi, Phys. Rev. Lett, {\bf 102} 210401 (2009).

\bibitem{refsug} C.~H\"{o}rhammer and H.~B\"{u}ttner, J.~Stat.~Phys {\bf 133}, 1161, (2008).

\bibitem{talknerheat} P.~Talkner, M.~Campisi and P.~Hanggi, J.~Stat.~Mech. P02025, (2009).

\bibitem{polish} W. Pusz, S. L. Woronowicz, Comm. Math. Phys. {\bf 58}, 273-290 (1978).

\bibitem{u1} M. Johanning, A. Braun, N. Timoney, V. Elman, W. Neuhauser, and Chr. Wunderlich, Phys.~Rev. Lett., {\bf 102}, 073004, (2009).

\bibitem{u2} N. Navon, S. Kotler, N. Akerman, Y. Glickman, I. Almog, and R. Ozeri, Phys. Rev. Lett., {\bf 111}, 073001, (2013).

\bibitem{u3} P.~Schindler {\it et al.}, New. J. Phys., {\bf 15}, 123012, (2013).

\bibitem{u4} F. Schmidt-Kaler, H. H\"affner, S. Gulde, M. Riebe, G. Lancaster, J. Eschner, C. Becher and R. Blatt, Europhys. Lett., {\bf 65}, 587,(2004).

\bibitem{kafri} D.~Kafri and S. Deffner, Phys.~Rev.~A. {\bf 86}, 044302 (2012).

\bibitem{vedral} V.~Vedral, J.~Phys.~A {\bf 45}, 272001, (2012).

\bibitem{rastegin} A.~Rastegin, J.~Stat.~Mech. P06016, (2013).

\bibitem{albash} T.~Albash, D.~A.~Lidar, M.~Marvian, P.~Zanardi, Phys.~Rev.~E. {\bf 88} 032146, (2013).

\bibitem{rastegin2} A.~Rastegi and K.~\.Zyczkowski, Phys. Rev. E {\bf 89}, 012127 (2014).

\bibitem{recent} J.~Goold and K.~Modi, arXiv:1407.4618, (2014). 

\bibitem{bound} J.~Goold, M.~Paternostro, and K.~Modi, arXiv:1402.4499 (2014).

\bibitem{latent} E. Mascarenhas, H. Bragança, R. Dorner, M. França Santos, V. Vedral, K. Modi, and J. Goold, Phys.~Rev.~E {\bf 89}, 062103, (2014).


\end{thebibliography}
\end{document}